  \documentclass[
showpacs, amsmath, amssymb, floatfix,twocolumn,
nofootinbib,wrapfloat byrevtex]{revtex4}

 \usepackage{pstricks,pstricks-add,pst-eps}
 \usepackage{amsmath,mathrsfs,graphicx,epstopdf,placeins,float}

 \newcommand{\beq}[1]{\begin{equation}\label{#1}}
 \newcommand{\eeq}{\end{equation}}
 \newcommand{\bea}[1]{\begin{eqnarray}\label{#1}}
 \newcommand{\eea}{\end{eqnarray}}
 \newcommand\figcaption{\def\@captype{figure}\caption}
 \newcommand\tabcaption{\def\@captype{table}\caption}

\begin{document}

 \title{Baryon Asymmetries in the Natural Inflation Model}
 \author{Nan Li\footnote{linan11122@tom.com} \& Ding-fang Zeng\footnote{dfzeng@bjut.edu.cn}}
 \affiliation{Theoretical Physics Division, College of Applied Sciences, Beijing University of Technology}
 \begin{abstract}
 A variation of Affleck-Dine mechanism was proposed to generate the observed baryon asymmetry in \cite{17}, in which the inflaton was assumed to be a complex scalar field with a weakly broken $U(1)$ symmetry, and the baryon asymmetry generation was easily unified with the stage of inflation and reheating. We adapt this mechanism to natural inflation scenarios and compare the results with those in chaotic inflation models. We compute the net particle number obtained at the end of inflation and transform it into net baryon number after reheatings. We observed that in natural inflation models,
the desired baryon-to-photon ratio can be achieved equally well as in chaotic models.
 \end{abstract}
 \maketitle

 \tableofcontents

\section{Introduction}

The last 35 years may be the most rapidly developing 35 years for cosmologists. With the help of quickly-increasing data from observations, peoples now have established the so called standard model of universe. According to this model, the early universe experience a very short time of inflation\cite{2,3}, after which matter and anti-matter begin to form simultaneously through reheatings. If nothing special happens, the amount of matter and anti-matter should be equal. But observations indicate that there are more matter than anti-matter in the universe, the so-called
baryon asymmetry. Quantitatively, this is parameterized by the baryon-to-photon ratio~$\eta$~, whose observation value reads:
\beq{}
\eta_{obs}\thickapprox6\times 10^{-10}.
\label{etaObservation}
\eeq
It is unreasonable to explain this asymmetry as the initial condition of universe evolution. Because after inflations, any pre-inflation particle's number density would be diluted to zero so any asymmetries between matter and anti-matter should be wiped out totally. So, to implement the observed the baryon asymmetry, one must invoke some mechanism to generate a net baryon number after the inflation.

In 1967, Sakharov \cite{4} came up with three conditions that processes which can produce the baryon asymmetry should satisfy:
\begin{itemize}
\item The process violate baryon charge conservation.
\item The process violate C and CP invariance.
\item The process should take place in a nonequilibrium thermodynamic state.
\end{itemize}
The first condition comes directly, the second is for the decay of the particles and antiparticles to produce different
numbers of baryons and anti-baryons. The third is mainly to prevent the inverse process from annihilate the baryon asymmetry.

Among large number of theories trying to describe the baryon asymmetry, the most interesting one  may be the Affleck-Dine mechanism\cite{5}, which uses scalar field dynamics to get a net baryon number. Their basic idea is, in a matter or a radiation dominated universe, introducing a complex scalar field with $U(1)$-symmetry broken self-interaction and letting the evolution of the scalar field to produce the desired baryon asymmetry. In reference \cite{6}, by associating with inflation scenario, Andrei Linde give a more physical realization for this mechanism.  While in reference \cite{7,8,9,10,11,12,13,14,15,16}, more interesting and important questions are discussed.

Both in Affleck-Dine's original work and A. Linde's improvements, the complex scalar field is identified with the classical squark-slepton scalar field or their avatar, neither of which has direct relevance with inflatons. But in a recent work \cite{17}, the Hertzberg and Karouby proposed the idea that the complex scalar field just be the inflaton field $\phi$. So the non-zero net $\phi$-particle number is generated just during latter stage of inflation. While at the end of inflation, by reheating process, the net $\phi$-particles decay into baryons, thus achieve the desired baryon asymmetry.

Hertzberg and Karouby illustrated their idea with the chaotic inflation scenario\cite{17.5}. However, ignited by the recent BICEP2 observation \cite{1}, more focus of the community is attracted to the ``natural inflation''\cite{18} scenario. The natural inflation model is favored by its ``naturalness'' in physical realizations. As is well known \cite{19}, to solve the horizon-, flatness- and other related questions, any successful slow-roll single-field inflation model must satisfy
\beq{}
\chi\equiv\Delta V/(\Delta\phi)^4\leq \mathcal{O}(10^{-6}\sim10^{-8}),
\eeq
where $\Delta V$ and $\Delta\phi$ are the change of potential and field respectively during the  inflation era. This small ratio of mass scales required is known as the fine-tuning problem in inflation.
It quantifies how flat the inflaton potential should be. In the natural inflation, the flatness of the potential is easily achieved by a shift symmetry under which $\phi\rightarrow\phi + \mathrm{constant}$. Our purpose in this paper is just to adapt Hertzberg and Karouby's idea to the natural inflation model.

This paper is organized as follows: In Section II we introduce the natural inflation model with complex scalar fields and illustrate the process of inflaton asymmetries generation, in Section III
we let the inflaton decay into baryons and give our final results on baryon asymmetries. Section IV is a summary of our work and some discussions.

\section{Net $\phi$-particle generations in natural inflation models}

Let us begin our investigation from the simplest complex scalar field inflaton models, whose action has the form
\beq{}
S=\int d^4 x\sqrt{-g}~\left[\frac{1}{16\pi G}\mathcal{R} +\frac{1}{2}\,|\partial\phi|^2-V(\phi,\phi^\ast)\right],\eeq
where $g$ is the determinant of the metric, $\mathcal{R}$ is Ricci scalar and $V(\phi,\phi^\ast)$ is the effective potential of inflaton field $\phi$.
In the current paper we will use the normal flat FRW metric with signatures (+ - - -) and natural units $\hbar=c=1$. Differences between various
inflation models root in their potential function $V$. In the original natural inflation model, the potential, or the lowest order approximation
of the potential is generally of the form:
\beq{}
V(\phi)=\Lambda^4(1\pm\mathrm{cos}(N\phi/f)),
\eeq
where $\phi$ is real scalar field; $f$ is the characteristic scale of global symmetry spontaneously breaking; $\Lambda$ is a lower scale associating with some explicit soft symmetry breaking;
the choice of the sign do not affect the physical results and we will choose the negative sign in this paper; the coefficient $N$ is always assumed to be equal to 1. For our purpose in this paper, we will take $\phi$ as a complex scalar field and modify the natural inflation potential and decompose it into two parts:
\beq{}
V(\phi,\phi^\ast)=V_s(|\phi|)+V_b(\phi,\phi^\ast).
\eeq
The $V_s$ part conserves the global $U(1)$ symmetry while the $V_b$  part breaks down it. To implement the desired natural inflation, we assume that $V_s$ always dominates and set
\beq{}
V_s(|\phi|)=\Lambda^4(1-\mathrm{cos}(\frac{|\phi|}{f})).
\eeq
Obviously $V_s(\phi)$ is invariant under the global $U(1)$ transformation $\phi\rightarrow e^{-i \alpha}\phi$. According to Noether's theorem, any global symmetry
leads to a conserving current. The $U(1)$ symmetry of $V(\phi)$ is related to the net $\phi$-particle number $N_{\phi}-N_{\bar{\phi}}$. So, to obtain a non-zero $\phi$-particle
number which is related to baryon numbers from an initially $\phi\bar{\phi}$ symmetric universe, we must break down this $U(1)$ symmetry. Following Hertzberg and Karouby, we implement this goal by setting the symmetry breaking part in the potential as:
\beq{}
V_b(\phi,\phi^\ast)=\lambda(\phi^n +\phi^{\ast n}),
\eeq
where the integer $n\geq 3$ and $\lambda$ is a symmetry breaking parameter. Although the cross terms like $\phi^{n-m}\phi^{\ast m}+\phi^{\ast n-m}\phi^m$ also break the $U(1)$ symmetry, we will not consider them for simplicities.

The smallness of $\lambda$ is natural in physics by 't Hooft's critirial \cite{20}: a small parameter in a theory is natural if, in the limit it is set to zero, the symmetry of the system increases. Obviously, when $\lambda=0$, the $U(1)$ symmetry is recovered, and
the symmetry of the system increases. We also need the smallness of $\lambda$ to preserve the shape of the potential for inflaton, otherwise the character of the natural inflation would be destructed. From the observation aspect, a small value of $\lambda$ is also favored by small baryon-to-photon ratios. Because $\lambda$ is just the measure of $U(1)$-symmetry breaking degree which is responsible for the net particle number's generation.

It's worth mention that despite the $U(1)$ symmetry is broken by $V_b$, the charge conjugation symmetry $\phi\leftrightarrow\phi^{\ast}$ is still respected. We assume that this symmetry is broken in the following process, or the Sakharov's conditions would be violated. In the original Affleck-Dine mechanism, it is spontaneously broken by the interaction with some other light fields. However, the detailed mechanism is
not important to us, and it do not affect our results, so we will not discuss it in this paper.

\subsection{Net $\phi$-particles from $\phi$ and $\bar{\phi}$}

Firstly, noting that the function $V_s$ is a periodic function of period $2\pi f$, we restrict
the value of $|\phi|\in[0,\pi f]$. Secondly, using the fact that $V_s$ takes minimum at $|\phi|=0$, we make Taylor expansion of it at this point as $V_s\approx \frac{1}{2}\Lambda^4(\frac{|\phi|}{f})^2$ when $|\phi|$ is small. Thirdly, since $n\geq3$ in $V_b$, at the late time of inflation during which $|\phi|$
is small, $V_b$ decreases faster than $V_s$ so soon becomes negligible. As results, the effective potential of the inflaton at later times conserves the global $U(1)$ symmetry. According to Noether's theorem, we can derive out the conserving charge as the net particle number:
\beq{}
\Delta N_\phi=N_\phi-N_{\bar{\phi}}=i\int d^3 x \sqrt{g_s} \,(\phi^\ast\, \dot{\phi}-\dot{\phi}^\ast\, \phi),
\eeq
here $d^3 x \sqrt{g_s}$ is the spatial volume measure, $N_\phi$ and $N_{\bar{\phi}}$ are the number of $\phi$-and $\bar{\phi}$-particles.
As the roughest approximation, we take $\phi$ as spatial-homogeneous. Substituting the FRW metric into this definition, we can work out the integral and get:
\beq{}
\Delta N_\phi=N_\phi-N_{\bar{\phi}}=iV_{com}a(t)^3\,(\phi^\ast\, \dot{\phi}-\dot{\phi}^\ast\, \phi),
\eeq
where $V_{com}$ is the comoving volume and $a(t)$ is the scale factor.

To get the equation of motion for $\phi$, we vary the total action of the system with respect to $\phi^\ast$ and get
\beq{}
\ddot{\phi}+3H\dot{\phi}+\frac{\Lambda^4}{f}\sqrt{\frac{\phi}{\phi^\ast}}\mathrm{sin}(\frac{|\phi|}{f})+2\,\lambda\, n\, \phi^{\ast n-1}=0,
\label{eomPhi}
\eeq
where $H=\dot{a}/a$ is the Hubble parameter. Taking the time derivative of $\Delta N_\phi$
\beq{}
\frac{\partial}{\partial t} \Delta N_\phi=i\, V_{com}a^3(3H(\phi^\ast\, \dot{\phi}-\dot{\phi}^\ast\, \phi)+\phi^\ast\, \ddot{\phi}-\ddot{\phi}^\ast\, \phi),
\eeq
and substituting the results into an appropriate combination of \eqref{eomPhi} with its complex conjugate, we will get
\bea{}
\Delta N_\phi(t_f)=\Delta N_\phi(t_i)+2i\,\lambda\,V_{com}\,n\,\int^{t_f}_{t_i}\,dt\,a(t)^3\times
\\
(\phi(t)^n-\phi^\ast(t)^n),
\nonumber
\eea
where $\Delta N_\phi(t_i)$ is the initial net particle number at time $t_i$, while $t_f$ denotes the final time. From this equation, we easily see that when the $U(1)$ symmetry is unbroken, i.e. $\lambda\rightarrow0$, the net particle number will indeed be conserved. Since any initial particle number would be diluted by inflation, and the process we feel interested in
happens at the late time of inflation, we will set $\Delta N_\phi(t_i)=0$ from now on.

For conveniences in the latter derivations, we express the scalar field $\phi$ in the polar coordinate as
\beq{}
\phi(t)=\Phi(t)\, e^{i\theta(t)},
\eeq
and rewrite the net particle number $\Delta N_\phi$ in the form
\beq{}
\Delta N_\phi(t_f)=-4\lambda\,V_{com}\,n \int^{t_f}_{t_i}\,dt\,a(t)^3\,\Phi(t)^n\,\mathrm{sin}(n\,\theta(t)).
\label{DeltaNphi}
\eeq
Like the field $\phi$ and $\phi^*$, the polar field $\Phi$ and angular field $\theta$ also satisfy differential equations similar to \eqref{eomPhi}, which can be solved order by order in $\lambda$.
By equation \eqref{DeltaNphi}, $\Delta N_\phi$ is proportional to $\lambda$. So if we need only calculate $\Delta N_\phi$ to first order approximation which is reasonable when $\lambda$ is small, then we only need to calculate
the integral to the zeroth order in $\lambda$. It can be proved that the evolution of $\theta$ is determined by the symmetry
breaking term. So when we neglect the effect of $\lambda$, $\theta$ doesn't evolve at all, i.e., $\dot{\theta}=0$. For this reason, the factor $\mathrm{sin}(n\,\theta(t))$ in (14) can be extracted out of the integrations. Using $\Phi_0(t)$ and $a_0(t)$ to denote $\Phi(t)$ and $a(t)$ when we
neglect the effect of $\lambda$ in the equations of motion, we can write $\Delta N_\phi$ as the form
\beq{}
\Delta N_\phi(t_f)=-4\lambda\,V_{com}\,n\, \mathrm{sin}(n\,\theta_i)\int^{t_f}_{t_i}\,dt\,a_0(t)^3\,\Phi_0(t)^n ,
\eeq
where $\theta_i$ is the initial value of $\theta$.\\

The equation of motion for $\Phi_0$ is easy to derive:
\beq{}
\ddot{\Phi}_0+3H_0\dot{\Phi}_0+\frac{\Lambda^4}{f}\mathrm{sin}(\frac{\Phi_0}{f})=0,
\eeq
while the corresponding Friedmann equation for $H_0$ reads
\beq{}
H^2_0=\frac{8\pi}{3 m^2_{Pl}}(\frac{1}{2}\dot{\Phi}_0^2+\Lambda^4(1-\mathrm{cos} (\frac{\Phi_0}{f}))),\eeq
where $m_{Pl}\equiv1/\sqrt{G}=1.22\times10^{19}$GeV is the Plank mass. By these two equations of motion, supplemented with appropriate initial conditions, we will be able to do the integral in eq.(15) very fluently.

\subsection{The value of $\Lambda$ and $f$ from observations}
According to reference \cite{24}, to be consistent with known cosmic-microwave background observations such as WMAP\cite{21}, the Planck\cite{22,23} and the BICEP2\cite{1}, the $\Lambda$ and $f$ parameters in the natural inflation should satisfy that $f\gtrsim m_{_{Pl}}$ and $\Lambda\thicksim m_{\mathrm{GUT}}\thicksim 10^{16}$GeV. Although the natural inflation in this paper is implemented with complex scalar fields, the parameter determination logic could be adapted from \cite{24} routinely.
\begin{itemize}
\item Constraints from the density perturbation spectrum index $n_s$. Both in real and complex scalar field, we can derive that
\beq{}
n_s=1-\frac{m^2_{Pl}}{8\pi f^2}.
\eeq
While according to reference \cite{23}, the observation value of $n_s\approx0.96$. This means that in the complex scalar field natural inflation model, $f\approx m_{_{Pl}}$
\item  Constraints from the tensor-to-scalar(perturbation amplitudes) ratio. Theoretical considerations \cite{24} require
\beq{}
V_H=(2.2\times 10^{16}\mathrm{GeV})^4\,\frac{r}{0.2},
\eeq
for natural inflation models $V_H=2\Lambda^4$. According to the observation of BICEP2 $r=0.20^{+0.07}_{-0.05}$. So $\Lambda\approx 10^{16}\mathrm{GeV}$ is very normal choices
\end{itemize}

According to reference \cite{25}, the number of inflation e-foldings to solve the flatness- and horizon-problem of non-inflation cosmologies also implies constraints on the choice of model parameters $\Lambda$ and $f$. Its basic logic is as follows.

Firstly, according to the slow-roll scenario, the number of inflation e-foldings reads
\bea{}
N_e=\ln(\frac{a_2}{a_1})=\int^{t_2}_{t_1}H\,dt=\frac{8\pi}{m_{Pl}^2}\int^{\phi_1}_{\phi_2}\frac{V(\phi)}{V'(\phi)}d\phi
\\
=\frac{8\pi f^2}{m_{Pl}^2}\ln\left[\frac{1+\cos(|\phi_2|/f)}{1+\cos(|\phi_1|/f)}\right],
\nonumber
\eea
where $a_1$ and $\phi_1$ are initial values when the inflation begins, $a_2$ and $\phi_2$ are the values at the end of inflation, and $V'$ denotes $dV/d\phi$.

\begin{center}
\begin{figure}[ht]
\includegraphics[width=7cm]{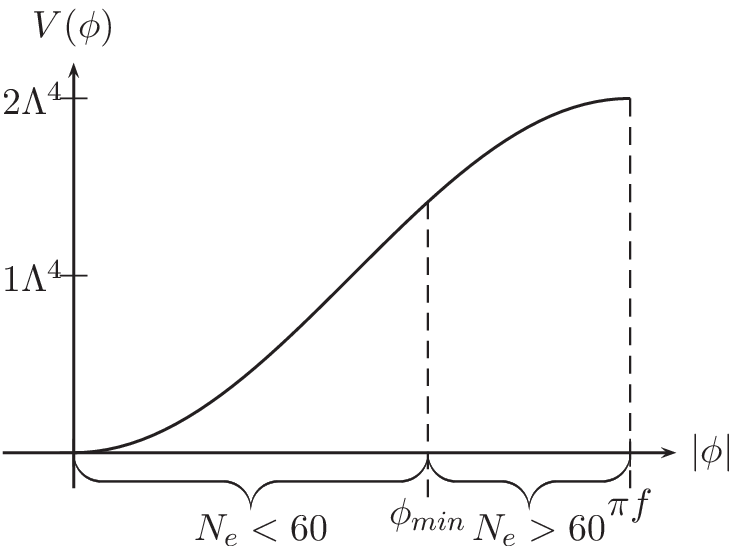}
\includegraphics[width=8cm]{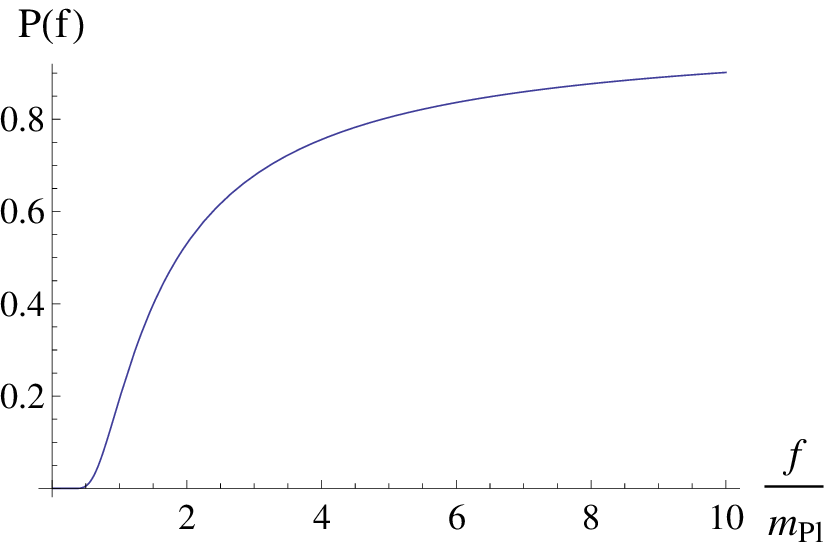}
\caption{The top figure illuminates the shape of $V(\phi)$. It's clear that a higher potential generates a larger e-folding number, so
$|\phi_1|\in(\phi_{min},\pi f)$ would get $N_e>60$. The button figure shows the numeric feature of $P(f)$, from which we easily see that the larger is $f$, the more closer $P(f)\rightarrow1$.}
\label{figVphiPf}
\end{figure}
\end{center}

Secondly, to associate $N_e$ with $f$, define a ``possibility'' $P(f)$ to quantify whether a given $f$ value is likely to generate sufficient inflation that $N_e\approx 60$,
\beq{}
P(f)=\frac{\pi f-\phi_{min}(f)}{\pi f},
\eeq
where $\phi_{min}(f)$ is the minimal value of $|\phi_1|$ that getting $N_e\geq60$ for a given $f$. Obviously, as long as $\phi_{min}$ can drive
sufficient inflation, all values of $|\phi_1|\in(\phi_{min},\pi f)$ will yield $N_e>60$, as shown in Fig \ref{figVphiPf}.

Thirdly, using the slow-roll parameter definition
\beq{}
\epsilon=\frac{m_{Pl}^2V'(\phi)^2}{16\pi V(\phi)^2}=\frac{m_{Pl}^2}{16\pi f^2}\left[\frac{\sin(\phi/f)}{1-\cos(\phi/f)}\right]^2,
\eeq
and the inflation ending condition $\epsilon\approx1$, we can get the $\phi$ field value at the inflation ending:
\beq{}
\phi_2(f)=f\,\arccos\left[\frac{16\pi f^2-m_{Pl}^2}{16\pi f^2+m_{Pl}^2}\right].
\eeq

Finally, combining eqs.(20)(21)(23) we can exactly work out $P(f)$ and the result is shown in Fig \ref{figVphiPf}. While from the definition of $P(f)$, we know that to get enough number of inflation e-foldings, $P(f)$ should be as close as possible to $1$. For $f=m_{Pl}$, we get $P(f)=0.194$, which is grudgingly in the desired range. More large values of $f$ will generate more sufficient inflation.

In following calculations, we will set $f=m_{Pl}$ and $\Lambda=10^{16}$GeV when necessary.

\subsection{Dimensionless representation}

\begin{center}
\begin{figure}[ht]
\includegraphics[width=7.5cm]{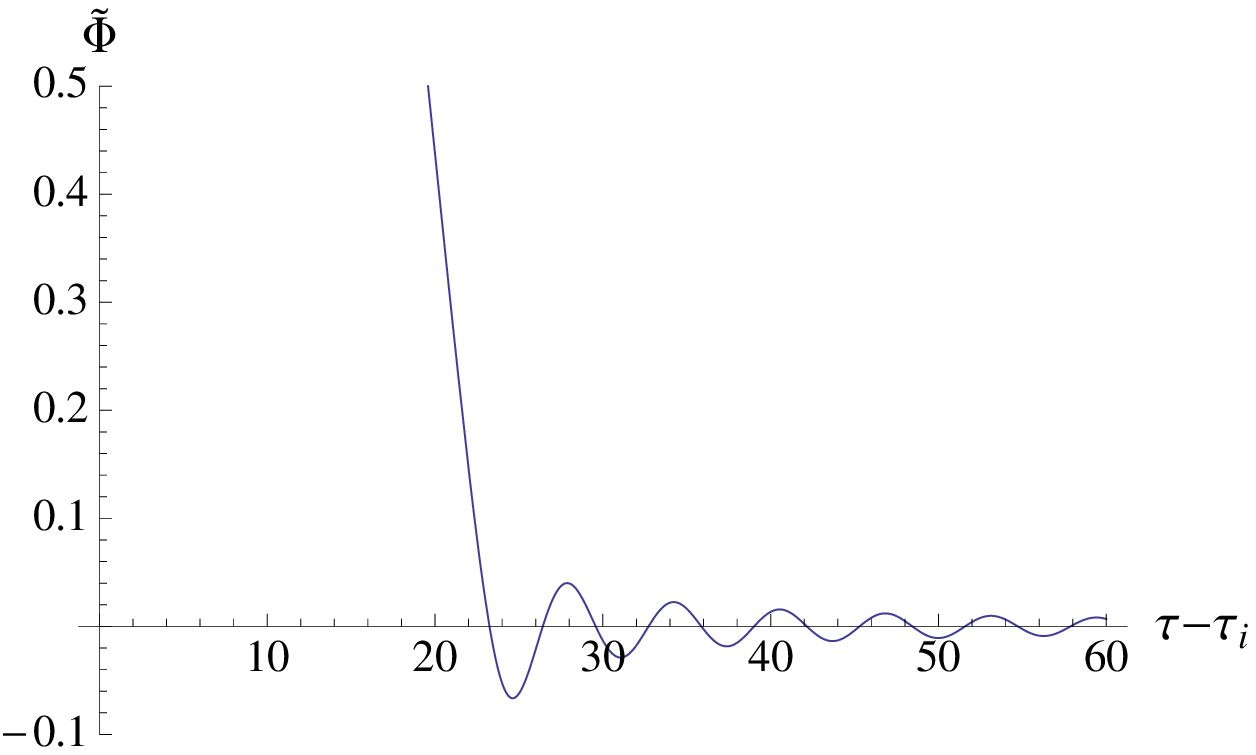}
\includegraphics[width=7.5cm]{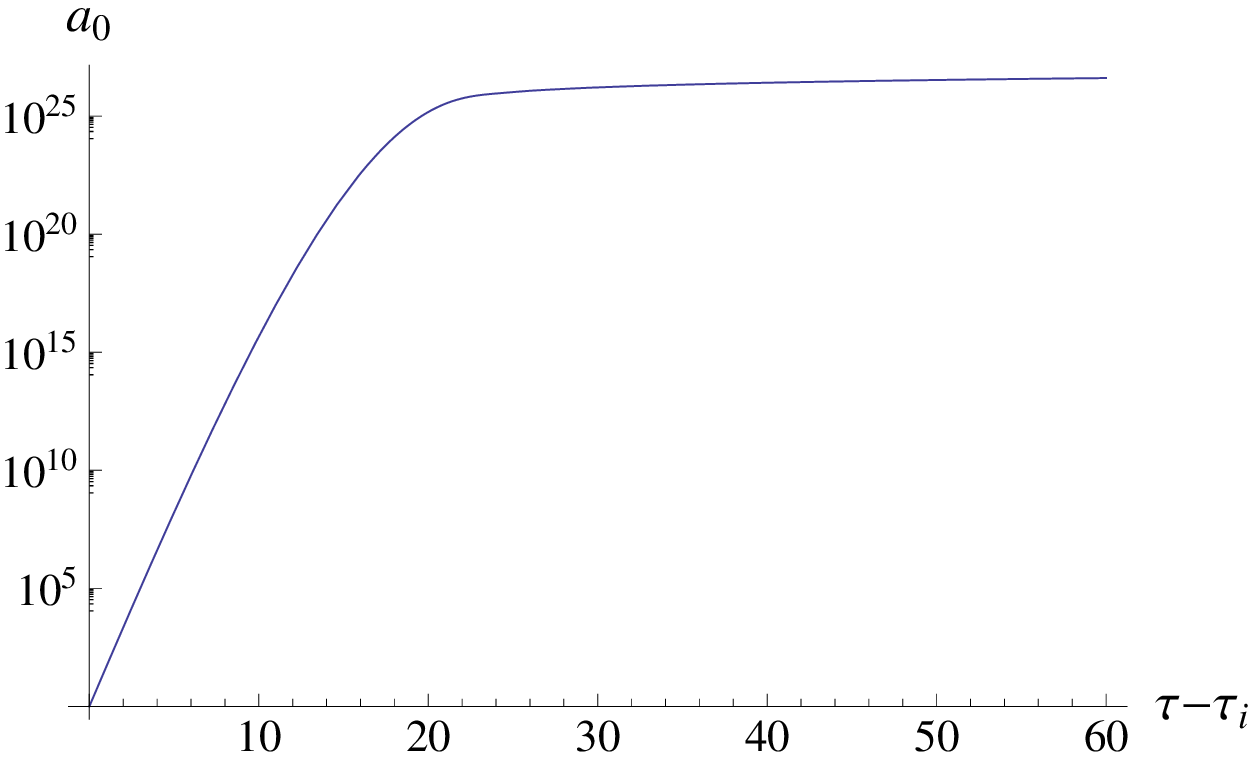}
\caption{
The evolution of the dimensionless field variable $\tilde{\Phi}$ and scale factor $a_0$.
In this plot, the initial conditions were set to $\tilde{\Phi}=2.5$, $f=m_{_{Pl}}$ and $\Lambda=10^{16}$GeV to implement the e-folding number $\sim$60. Without loss of generality, we set $\dot{\tilde{\Phi}}=0$ and $a_i=1$. The right figure shows that indeed about 60 e-folding numbers are generated.}
\label{figphita0}
\end{figure}
\end{center}

Since $\Delta N_\phi$ is proportional to the size of the expanding universe, it is not a good quantity for numerics, even though it is dimensionless. The more appropriate quantity measuring the baryon asymmetry is
\beq{}
\alpha\equiv \frac{\Delta N_\phi}{N_{tot}}=\frac{\Delta n_\phi}{n_\phi+n_{\bar{\phi}}},
\eeq
where $N_{tot}$ is the total number of $\phi$ and $\bar{\phi}$ particles and $n=N/V_{com}a^3$ stands for particle number densities. After the inflation finishes, but before the decay of $\phi$ particles into baryons, all energies that fill the universe is stored in the non-relativistic $\phi$ particles. So we have
\beq{}
m_\phi(n_\phi+n_{\bar{\phi}})=\varepsilon_0=\frac{1}{2}\dot{\Phi}_0^2+\Lambda^4(1-\mathrm{cos} (\frac{\Phi_0}{f})),
\eeq
where $m_\phi=\frac{\Lambda^2}{f}$ is the mass of $\phi$ particle and $\varepsilon_0$ is the energy density of the universe. From the above two equations, we can derive
\beq{}
\alpha=\frac{m_\phi \Delta n_\phi}{\varepsilon_0} .
\eeq
To get further dimensionless representation for $\alpha$, we introduce the following dimensionless quantities:
\beq{}
\tau\equiv m_\phi t=\frac{\Lambda^2 t}{f},~~\tilde{\Phi}\equiv \frac{\Phi_0}{f},~~\tilde{H}\equiv\frac{H_0}{m_\phi}=\frac{fH_0}{\Lambda^2},
\eeq
and write
\beq{}
\alpha=-\frac{\lambda\,f^n}{\Lambda^4}\,\mathrm{sin}(n\,\theta_i)\,A_n(\tau_i,\tau_f),
\label{alphabyAn}
\eeq
where $\tau$ and $\tilde{\Phi}$ are dimensionless time and field variables respectively, while
\beq{}
A_n(\tau_i,\tau_f)=\frac{4\,n\,\int^{\tau_f}_{\tau_i}d\tau\,a_0(\tau)^3\tilde{\Phi}(\tau)^n}{a_0(\tau_f)^3(\frac{1}{2}\dot{\tilde{\Phi}}(\tau_f)^2+1-\mathrm{cos}(\tilde{\Phi}(\tau_f)))}.
\label{Antitf}
\eeq
By numerically solving the dimensionless version of equations (16) and (17), we will obtain the time dependence of $\tilde{\Phi}_0(\tau)$, $a_0(\tau)$ very easily, see FIG \ref{figphita0} for references.

\begin{center}
\begin{figure}[ht]
\includegraphics[width=7.5cm]{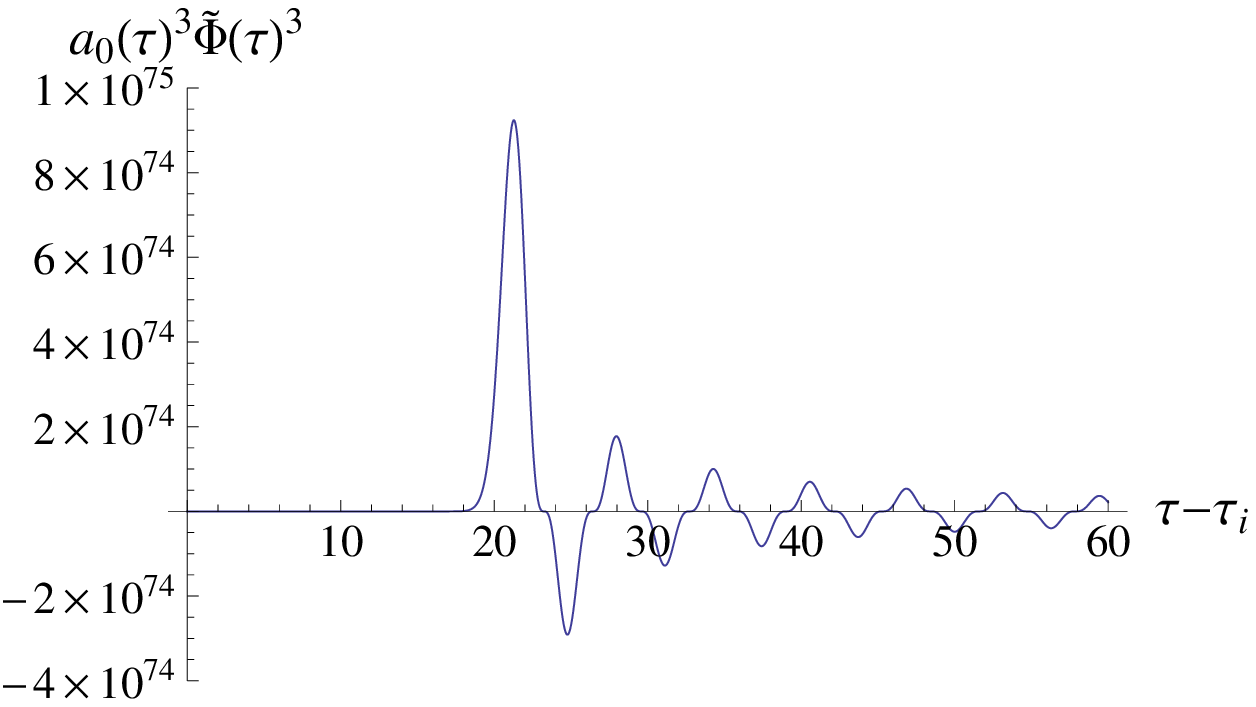}
\includegraphics[width=7.5cm]{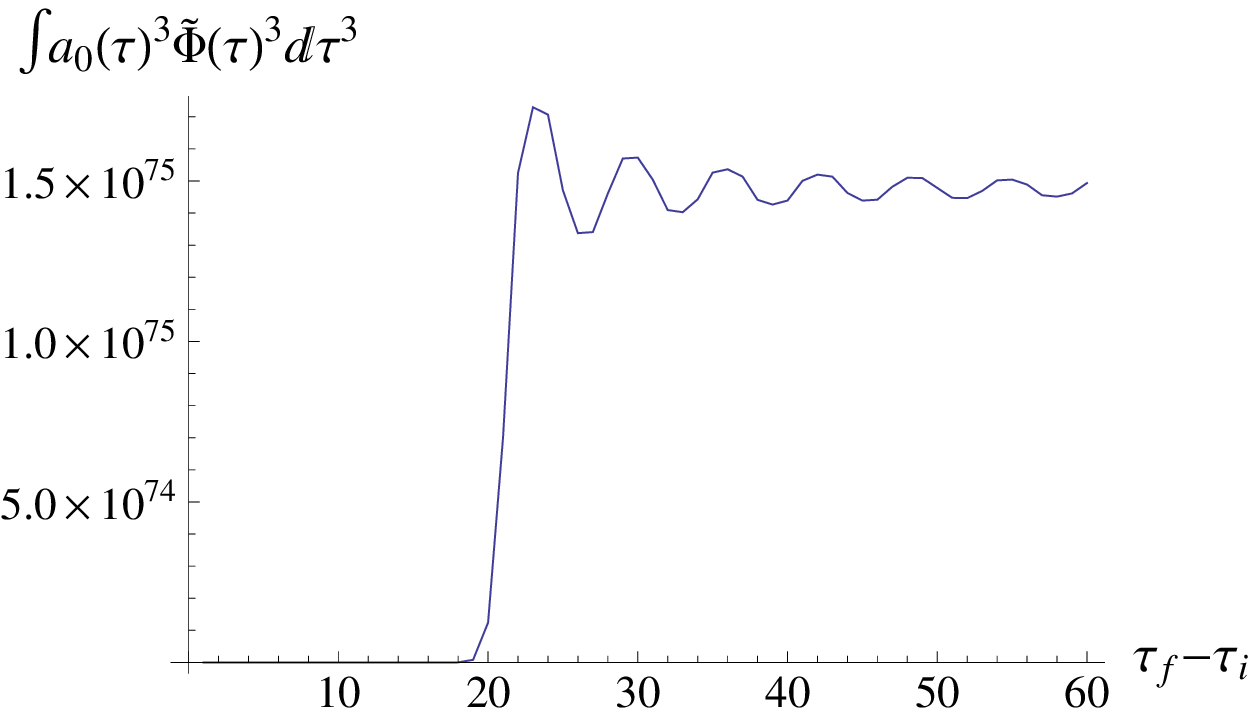}
\caption{
The top is the integrand in eq.(29), which is proportional to the production rate of the net particle number. The button is the integrated value, expresses the whole net particle number produced till $\tau_f$. We can easily see that the early and late time contributions do not significantly affect the final net particle number. Almost all of the net particles are produced during a very short time around $\tau=20$.
}\label{figa3phi3related}
\end{figure}
\end{center}

From FIG \ref{figphita0}, we can easily see that at the beginning of inflation,  $a_0(\tau_i)$ is very small(relative to $a_0(t_f)$). As results, in the integration (29) contributions from the early time are negligible. While at the matter dominating era marked by $\tau_f$, $\tilde{\Phi}$ is evolving to almost zero. So integrations from that period also contribute little to $A_n$, see FIG \ref{figa3phi3related} for quantitative references. From the figure, it's easy to see that $A_n$ is totally determined by the ``middle'' area of the integrand. Under the limit that $\tau_i\rightarrow 0$ and $\tau_f\rightarrow \infty$, $A_n$ can be looked as a constant which depends only on the lower index $n$;
\beq{}
A_n(\tau_i\rightarrow 0,\tau_f\rightarrow\infty)\equiv c_n.
\eeq
As examples, we numerically compute this parameter when $n=3,4,\cdots,10$, the result is as follows
\beq{}
\begin{array}{c}
c_3\approx8.0,~c_4\approx3.7,~c_5\approx1.3,~c_6\approx0.56
\\
c_7\approx0.25,~c_8\approx0.12,~c_9\approx0.060,~c_{10}\approx0.031
\end{array}
\eeq
Substituting this results into equations \eqref{alphabyAn}-\eqref{Antitf}, we will get the final expression for the dimensionless baryon asymmetry parameter as following
\beq{}
\alpha=-c_n\,\frac{\lambda\,f^n}{\Lambda^4}\,\mathrm{sin}(n\,\theta_i).
\label{alphaFourth}
\eeq
Obviously, for some special values of the initial angle $\theta_i$, for
instance $\theta_i=\pi$, the factor $\sin(n\,\theta_i)$ vanishes. In such cases, no baryon asymmetry is generated. Such special values can generate
large isocurvature fluctuations \cite{17}. But that's not our main goal, we will set $\theta_i$ to be general values so that $|\sin(n\,\theta_i)|\approx 1$.\\

Now we have implemented the goal of generating net inflaton $\phi$-particles from a symmetric initial conditions in natural inflations. Our next goal is transferring the $\phi$-particles into baryons, and associating $\alpha$ to the observable baryon-to-photon ratio $\eta$ in the next section.

\section{$\phi$-particles decay into baryons}

According to inflationary theory, at the late time of inflation, the inflaton field oscillates near the minimum of its effective potential and gradually
decays into standard model particles\cite{27}. This stage of the early universe is called ``reheating''. Almost all elementary particles populating
the universe are created during reheating, and these particles interact with each other and finally come to a state of thermal
equilibrium at a temperature $T_r$, which is called reheating temperature.

Now we assume that each $\phi$ particle carries a baryon number $B$, and it will decay into baryons through a process that conserves the
baryon number during the stage of reheating.  We assume that all the
subsequent interactions also conserve the baryon number, so we have
\beq{}
(N_b-N_{\bar{b}})_f=B(N_\phi-N_{\bar{\phi}})_i,
\eeq
where the lower index $f$ means a final time on which reheating finishes, and $i$ stands for an initial time on which all the energy stored in the inflaton field is translated into $\phi$ particles, but $\phi$'s decay does not begin. By these symbols,  we can write down $\eta$ in the following form:
\beq{}
\eta=\frac{(N_b-N_{\bar{b}})_f}{(N_{\gamma})_f}=B\frac{(N_\phi-N_{\bar{\phi}})_i}{(N_{\gamma})_f}=\alpha B\frac{(N_\phi+N_{\bar{\phi}})_i}{(N_{\gamma})_f}.
\label{etaFirst}
\eeq
Obviously, to calculate $\eta$, we need to work out the initial total number of $\phi$ particles and the photon number at late times.

At initial times, all the energy congesting the universe is provided by $\phi$ particles, so
\beq{}
m_\phi(N_\phi+N_{\bar{\phi}})_i=\frac{\Lambda^2}{f}(N_\phi+N_{\bar{\phi}})_i=V_{com}(a^3\varepsilon)_i .
\eeq
Using Friedmann equation, we can relate the energy density to the Hubble parameter as
\beq{}
(\varepsilon)_i=\frac{3m_{Pl}^2}{8\pi}(H^2)_i .
\eeq
While on the number of photons at late times, we can relate it with the temperature:
\beq{}
(N_\gamma)_f=V_{com}(a^3 n_\gamma)_f=V_{com}\frac{2\zeta(3)}{\pi^2}(a^3T^3)_f ,
\eeq
where $\zeta(3)\approx1.202$ is the so called Ap\'{e}ry's constant. Using this two results, we can rewrite the ratio $\eta$ in equation \eqref{etaFirst} as follows:
\beq{}
\eta=\frac{3\pi B\alpha}{16\zeta(3)}\frac{m_{Pl}^2 f}{\Lambda^2}\frac{(a^3H^2)_i}{(a^3T^3)_f}.
\label{etaSecond}
\eeq
Since we are not going to compute the result by detailed decaying processes,
we need to assume the decay of $\phi$ particles and the subsequent process of thermalization occurs very fast. So we can set both $(a^3H^2)_i$ and $(a^3T^3)_f$ to be values around the end of reheating and get an approximation for the final result. We insert an $\mathcal{O}(1)$ factor $\beta$ to account for the deviation caused by this assumption:
\beq{}
\eta=\frac{3\beta\pi B\alpha}{16\zeta(3)}\frac{m_{Pl}^2 f}{\Lambda^2}\frac{H^2_r}{T^3_r}.
\label{etaThird}
\eeq

The stage of reheating ends when Hubble parameter becomes smaller than the decay rate of $\phi$,
$H\lesssim \Gamma_\phi$\cite{27}. And the reheating temperature can be estimated by\cite{26}: $T_r\approx0.2\sqrt{\Gamma_\phi m_{Pl}}$.
Substituting these two relations to the approximate expression of $\eta$, we obtain:
\beq{}
\eta=\frac{3\pi\beta B\alpha}{0.2^3\times16\zeta(3)}\frac{m_{Pl}^{1/2} \Gamma_\phi^{1/2} f}{\Lambda^2}.
\label{etaFourth}
\eeq
Inserting the $\alpha$ expression \eqref{alphaFourth} obtained in the previous section into this equation, we will get our result for $\eta$:
\beq{}
\eta=-c_n\frac{3\pi\beta B\lambda}{0.2^3\times16\zeta(3)}\frac{m_{Pl}^{1/2} \Gamma_\phi^{1/2} f^{n+1}}{\Lambda^6}\sin(n\theta_i).
\label{etaFifth}
\eeq

Before further discussion of physical features of this expression for $\eta$, we should first determine the range of the symmetry breaking parameter $\lambda$. Since we assumed that the symmetry breaking term in the potential of $\phi$ is subdominant during the
inflation to ensure the feature of the natural inflation, we have to impose constraint
\beq{}
\lambda(\phi^n_i+\phi^{\ast n}_i)=\lambda \Phi^n_i \cos(n\,\theta_i)\ll \Lambda^4(1-\cos(\frac{\Phi_i}{f})),
\eeq
While to assure this constraint holds for all the possible value of $\theta_i$, the value of $\lambda$ has to be limited from the upper bound
\beq{}
\lambda\ll\lambda_{max}=\frac{\Lambda^4(1-\cos({\Phi_i/f}))}{\Phi_i^n}.
\eeq
With this constraint, we can test whether our result is physical acceptable.

Now we use the boundary value of $\lambda$ to work out the required $\Gamma_\phi$ for generating the observed $\eta\approx 6\times 10^{-10}$.
The expression of $\Gamma_{\phi,\mathrm{req}}$ can be derived from eq.(41):
\bea{}
\Gamma_{\phi, \mathrm{req}}\approx c_n^{-2}(\frac{\lambda_{max}}{\lambda})^2\frac{\eta^2(0.2^3\times16\zeta(3)^2)}{(3\pi)^2}\times
\nonumber\\
\frac{\Lambda^4f^{-2n-2}\Phi^{2n}_i}{m_{Pl}(1-\cos(\frac{\Phi_i}{f}))^2}(\beta B|\sin(n\theta_i)|)^{-2}.
\eea
In the previous sections, we have set $f=m_{Pl}$, $\Lambda=10^{16} \mathrm{GeV}$ and $\Phi_i=2.5 f$, so the expression can be simplified to:
\beq{}
\Gamma_{\phi, \mathrm{req}}\approx1.6\times 10^{-7} \mathrm{eV}\times 2.5^{2n}c_n^{-2}(\frac{\lambda_{max}}{\lambda})^2(\beta B|\sin(n\theta_i)|)^{-2}.
\label{GammaphiFirst}
\eeq

If we set an appropriate value for $\lambda$, for example, $\lambda=\frac{1}{10}\lambda_{max}$, and assume that $\beta B |\sin(n\theta_i)|\approx1$,
we will get the required decay rate for different $n$s. Our results, and those from \cite{17} with a magnitude correction for comparisons, are shown as follows:
\begin{eqnarray*}
\raisebox{3mm}{n~~~~} &\stackrel{\displaystyle\Gamma_{\phi,\mathrm{req}}~\mathrm{in}}{\displaystyle\mathrm{natural~inflation}} & \stackrel{\Gamma_{\phi,\mathrm{req}}~\mathrm{in}}{\mathrm{chaotic~inflation}}\\
n=3  &  6.1\times 10^{-5}\mathrm{eV} & ~~~~~~~~4\times 10^{-3}\mathrm{eV}\\
n=4  &  1.8\times 10^{-3}\mathrm{eV} & ~~~~~~~~2\times 10^{-1}\mathrm{eV}\\
n=5   &  9.0\times 10^{-2}\mathrm{eV} & ~~~~~~~~10^{1}\mathrm{eV}\\
n=6   &  3.0 ~\mathrm{eV} & ~~~~~~~~6\times10^2~\mathrm{eV}\\
n=7   &  9.5\times10 ~\mathrm{eV} & ~~~~~~~~2\times 10^4 \mathrm{eV}\\
n=8   &  2.6\times 10^3\mathrm{eV}  & ~~~~~~~~9\times 10^5 \mathrm{eV}\\
n=9   &  6.5\times 10^4\mathrm{eV}  & ~~~~~~~~3\times 10^7 \mathrm{eV}\\
n=10  &  1.5\times 10^6\mathrm{eV} & ~~~~~~~~10^9 \mathrm{eV}
\end{eqnarray*}
Obviously, larger power $n$ of symmetry breaking interaction requires larger decay width to give desired photon-baryon-ratio. While from derivations \eqref{etaFifth}-\eqref{GammaphiFirst}, we know that $\Gamma_\phi\propto\Lambda^4$, that is, higher energy scale of inflation requires larger width of inflaton decays, otherwise the theoretical photon-baryon-ratio will deviate remarkably from expectations.

With these results of $\Gamma_{\phi,\mathrm{req}}$, we can work out the corresponding reheating temperature by the
relation $T_r\approx 0.2\sqrt{\Gamma_\phi m_{Pl}}$. An important
condition is that the reheating temperature must be higher than the typical temperature of big bang nucleosynthesis $\sim$MeV.
For the lowest value of $\Gamma_{\phi,\mathrm{req}}$ in natural inflation, when $n=3$, the reheating temperature $T_r\approx 173$GeV, and for $n=10$, the corresponding $T_r\approx 2.7\times 10^7$GeV. All the $T_r$s in our model are much higher than MeV, this is obviously consistent
with the big bang nucleosynthesis, thus physically acceptable.

For all values of $n$ listed above, $\Gamma_{\phi,\mathrm{req}}$ in natural inflations is smaller than that in chaotic inflations, and the growth of $\Gamma_{\phi,\mathrm{req}}$ with the increasing of $n$ is slower than in chaotic inflation. These differences may be used to distinguish this two models in futures.

\section{Summary and discussion}

In this paper, we apply a variation of Affleck-Dine mechanism into natural inflation scenarios and generate the observed baryon-to-photon ratio. In this mechanism, the process of baryon asymmetry generation is unified with the stage of inflation and reheating. The baryon asymmetry is firstly implemented using the inflaton field with a weakly broken global $U(1)$ symmetry. It is in the second stage that the net inflaton $\phi$-particles decay into standard model particles.

By numerical calculations, we work out parameter $\alpha$ describing the asymmetric evolution of $\phi-\bar{\phi}$ particles during the natural inflation era and derive out formulas relating it with the baryon-to-photon ratio $\eta$. We calculate the decaying rate of $\phi$-particles required to generate the observed $\eta\approx 6\times10^{10}$. It is observed that
the reheating temperatures in this inflation model is much higher than the desired temperature of big bang nucleosynthesis. From this aspect, this model is physically acceptable. We also compare our results with those in chaotic inflation models. The differences between the two may be useful
for future distinguishing of them through observations.

As discussions, we note that parameter resonance phenomena\cite{26}, superheavy fermions production\cite{28}, detailed particle physics model implementation, dark matter particles formation and properties \textit{et al}. in this natural inflation + reheating mechanism are all interesting future directions.

\section*{Acknowledgments}

We thank very much to Wang Qian-jun, Meng Sun, Jian-feng Wu and Prof. Yong-chang Huang for meaningful discussions. This work is supported by Beijing Municipal Natural Science Foundation, Grant. No. Z2006015201001.

 \end{document}